\documentclass{osa-article}

\DeclareTextSymbol{\degre}{T1}{6}

\journal{osac}

\begin{document}

\title{Analytical decomposition of Zernike and hexagonal modes over an hexagonal segmented optical aperture}

\author{Pierre Janin-Potiron,\authormark{1,2,*} Patrice Martinez,\authormark{3} and Marcel Carbillet\authormark{3}}

\address{\authormark{1}ONERA The French Aerospace Laboratory, F-92322 Ch\^atillon, France\\
\authormark{2}Aix Marseille Universit\'e, CNRS, CNES, Laboratoire d'Astrophysique de Marseille (LAM), UMR 7326, 13388 Marseille, France\\
\authormark{3}Universit\'e C\^ote d'Azur, Observatoire de la C\^ote d'Azur, CNRS, Laboratoire Lagrange, UMR 7293, CS 34229, 06304 Nice Cedex 4, France}

\email{\authormark{*}pierre.janin-potiron@lam.fr} 



\begin{abstract}
Zernike polynomials are widely used to describe common optical aberrations of a wavefront as they are well suited to the circular geometry of various optical apertures. Non-conventional optical systems, such as future large optical telescopes with highly segmented primary mirrors or advanced wavefront control devices using segmented mirror membrane facesheets, exhibit an hexagonal geometry, making the hexagonal orthogonal polynomials a valued basis.
Cost-benefit trade-off study for deriving practical upper limit in e.g., polishing, phasing, alignment, and stability of hexagons imposes analytical calculation to avoid time-consuming end-to-end simulations, and for the sake of exactness.  Zernike decomposition over an hexagonal segmented optical aperture is important to include global modes over the pupil into the error budget. However, numerically calculated Zernike decomposition is not optimal due to the discontinuities at the segment boundaries that result in imperfect hexagon sampling.
In this paper, we present a novel approach for a rigorous Zernike and hexagonal modes decomposition adapted to hexagonal segmented pupils by means of analytical calculations.
By contrast to numerical approaches that are dependent on the sampling of the segment, the decomposition expressed analytically only relies on the number and positions of segments comprising the pupil. Our analytical method allows extremely quick results minimizing computational and memory costs. 
Further, the proposed formul\ae~can be applied independently from the geometrical architecture of segmented optical apertures. Consequently, the method is universal and versatile per se. 
For instance, this work finds applications in optical metrology and active correction of phase aberrations. 
In modern astronomy with extremely large telescopes, it can contribute to sophisticated analytical specification of the contrast in the focal plane in presence of aberrations.
\end{abstract}

\section{Introduction}

The Zernike polynomials (e.g., \cite{Born99, Mahajan13}) are commonly used in various fields of optics because they offer an orthonormal basis representing balanced classical optical aberrations defined on the unit disk over which  the phase of wavefronts can be decomposed in a unique way.
Because they are well adapted to the circular shape of most of the conventional optical systems, the Zernike polynomials became the standard way to describe optical path differences in wavefronts in many fields ranging from precision optical design and testing \cite{Malacara}, atmospheric and adaptive optics \cite{Noll76, Roddier90}, optical cophasing \cite{Pinna2007}, vision science \cite{Manzanera11,Liang97}, or optical communications \cite{Levine98}. 

Even though the circle is the most common optical aperture shape, they are other important geometries, such as the hexagonal structure. For instance, the next generation of ground-based astronomy imaging telescopes (e.g., the TMT or the ELT projects of more than 30 m diameter \cite{Tamai2016,Nelson2006}) will exhibit a segmented primary mirror made of several hundreds of individual hexagonal segments.
Control authority for the wavefront is largely assigned to a second stage of optical system and thus on active or adaptive optics element (i.e., small-size deformable mirrors.)
In adaptive optics, advanced wavefront control devices such as small-size segmented deformable mirrors \cite{Helmbrecht2006, Helmbrecht10, Helmbrecht11, Helmbrecht16} made of hexagonal segments 
are intensively used in astronomy \cite{NDiaye13, Pope14, Martinez14}, laser shaping \cite{Norton14}, retinal imaging systems \cite{Miller05, Manzanera11}, microscopy \cite{Sinefeld15, Blochet17} or laser communication \cite{Baker04}. 

The importance of the hexagonal geometry led to the development of orthonormal polynomials on the unit hexagon \cite{Mahajan07,Mahajan06} and to 
ad hoc solutions for decomposing Zernike or hexagonal modes adapted to hexagonal segmented optical apertures.

The reproduction of an orthonormal basis onto the aperture of a segmented telescope is essential in order to provide a unique way of decomposing the phase in the aperture plan. Such a decomposition has applications in various domain of optics. It can be used to perform a modal control of an adaptive optic system or to allow high-performance open-loop correction of optical aberrations in a segmented pupil environment.
A recent study presented in~\cite{Leboulleux18} proposes to analyze the performance of a coronagraphic system under the presence of local piston and tip-tilt on each of the segments. Including in this formalism an analytical formulation of global orthonormal modes over a segmented pupil would allow a more precise description of the effects at stake in this process.
In this context, the needs concerning the generation of such basis using only the three degrees of freedom of each segment are many. We need to that end: (1) a versatile method that can provide the decomposition coefficients for any hexagonal segmented aperture; (2) a fast and reliable algorithm; (3) a method providing analytical results that can be used as is for theoretical calculation without loss of generality.

To our knowledge, only numerical methods have been proposed to reproduce a set of Zernike or hexagonal modes for a segmented pupil \cite{PrivateIris}. The piston, tip, and tilt to apply to each segment is generally obtained either by using a numerical projection of the signal on the first three Zernike modes, or by fitting a plane using a least square minimization. The main drawbacks are that the precision depends on the sampling of the hexagons used to reproduce the mirror, and scaling the process with the number of segments is precluded, meaning that the computational time-consuming calculations must be reprocessed anytime the configuration of the pupil changes.

In this context, we propose a general analytical framework to provide a rigorous Zernike and hexagonal modes decomposition adapted to hexagonal segmented pupils. 
We provide a unique general solution for the segment piston, tip, and tilt coefficients in order to reproduce accurately an arbitrarily set of Zernike or hexagonal modes. This solution, which naturally scales with the size, positions, rotation and number of segments comprising the mirror, is versatile per se.
On the contrary to numerical solutions, the analytical expressions are independent from the physical size of the pupil and can be applied on small deformable mirrors \cite{Helmbrecht2006, Helmbrecht10} as well as on future large optical telescopes \cite{Tamai2016,Nelson2006} without loss of precision in the decomposition..
While numerical methods require computational and memory cost which can be impressive, our analytical solution allows extremely quick results whatever the pupil complexity.
Finally, the resultant decomposition is mathematically exact which means it can be used as is in analytical calculation implying modal analysis in the presence of a segmented mirror.

In Section~\ref{sec:othornomal_polynommials} we present the rationale and the analytical expression of the Zernike and hexagonal decompositions, and in Section~\ref{sec:fitting_error} we discuss and evaluate the method.

\section{Analytical expressions of the Zernike and hexagonal decompositions}
\label{sec:othornomal_polynommials}

For pedagogical reasons and because it is considered as a basis for this work, we start by presenting the orthonormal polynomials on the unit hexagon. 
A set of orthonormal polynomials on the unit hexagon has been proposed  using the Gram-Schmidt algorithm \cite{Mahajan07,Mahajan06}. These polynomials are denoted $\mathrm{H}_j(x,y)$, where $x$ and $y$ stand for the coordinates in the pupil plane and $j$ corresponds to the Noll index notation \cite{Noll76} where the Zernike polynomials are sorted by increasing radial order and azimuthal frequency. The cosine polynomials are given an even $j$ and the sine polynomials an odd $j$. They can be expressed as a linear combination of the Zernike polynomials $\mathrm{Z}_j$ such as:
\begin{equation}
\mathrm{H}_j(x,y) = \sum\limits_{i=1}^{j} \alpha_i Z_i(x,y),
\end{equation}
where the coefficients $\alpha_i$ can be found in \cite{Mahajan07}.
These polynomials provide a unique decomposition of a wavefront over an hexagon. In the following, we use them as the basis for the decomposition. In practical cases, tip and tilt are usually set in units of radian. The correspondence between their values expressed in terms of the hexagonal polynomial coefficients, denoted $t$ and $T$, and the angle on the segment, denoted $\theta$, is given by $\theta = \arctan[2 \sqrt{\frac{6}{5}} t]$ for tip and $\theta = \arctan[2 \sqrt{\frac{6}{5}} T]$ for tilt.


Now, we address the problems of the projection of the Zernike and hexagonal modes over an hexagonal segmented pupil. Our objective is to calculate the coefficients in piston, tip, and tilt to apply to each segment to reproduce a given Zernike mode $\mathrm{Z}_j$, or a given hexagonal mode $\mathrm{H}_j$, over a segmented pupil.

\begin{figure}[!tbp]
\centering
{\includegraphics[width=7cm]{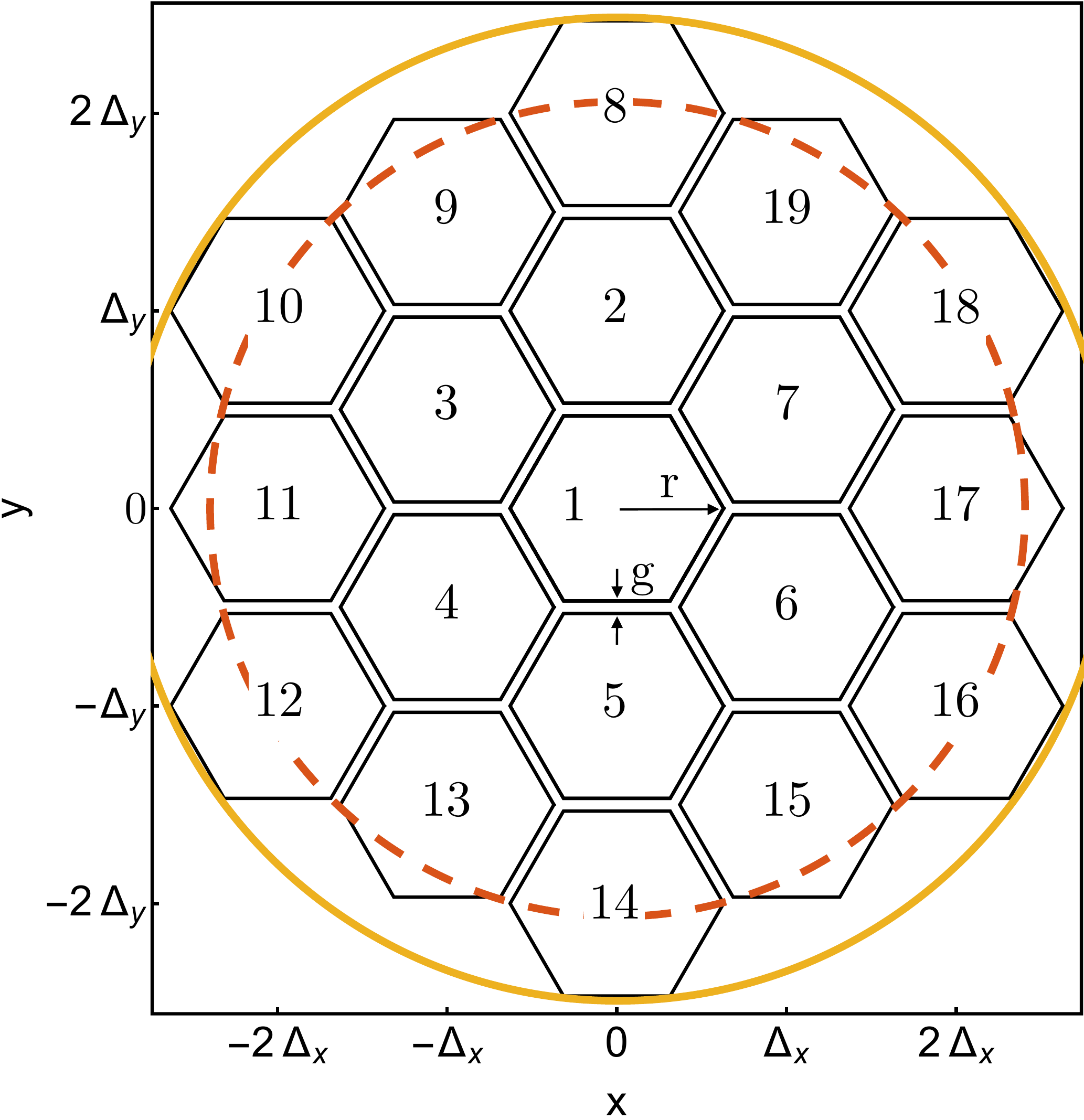}}
\caption{Representation of a $M=2$ ($N=19$) segmented pupil. The dashed red and solid orange circles stand for the inscribed and circumscribed circle to the full mirror respectively.}
\label{fig:segmented-pupil-variables}
\end{figure}

A segmented pupil is composed of N segments distributed over M rings around the central segment, where $N=3\times M \times (M+1) + 1$. Each segment can be actuated in piston, displacement along the z-axis; in tip, rotation around the $y$-axis; and in tilt, rotation around the $x$-axis. Figure~\ref{fig:segmented-pupil-variables} presents the variables introduced above for a $M=2$~segmented pupil.
We denote by $R$ the radius of the disk (or hexagon) on which the Zernike (or hexagonal) mode is applied.
We decide in the following of the paper to work with the disk circumscribed to the full segmented pupil (solid orange circle in Fig.~\ref{fig:segmented-pupil-variables}) in order to completely sample the mode. However, any other value of $R$ can be chosen so that the value is considered as a variable during the analytical calculations. Nevertheless, one has to keep in mind that for a diameter smaller than the circumscribed configurations, the decomposition at the segment boundaries is affected. This effect is out of scope and therefore not taken into account in this study.
We also note that the radius of the circumscribed circle is related to the number $M$ of rings comprising the pupil, the size $r_n$ center-to-corner of the single segment $n$ and the distance $g$ separating the segments. Its complete expression is given by $R=\frac{1}{2}\sqrt{r_n^2\left[1+3\times(2M+1)^2\right] + r_n \left[ 4\sqrt{3} M g (2M+1) \right] +(2Mg)^2}$. The area of the single segment $n$, denoted $A_n$ hereafter, is given by $A_n = 3\sqrt{3}r_n^2/2$.

The piston, tip, and tilt coefficients of the decomposition are obtained by projecting the Zernike mode $\mathrm{Z}_j$ respectively onto the hexagonal polynomials $\mathrm{H}_1$ for piston; $\mathrm{H}_2$ for tip; and $\mathrm{H}_3$ for tilt. They are expressed as:

\begin{equation}
p_n(\mathrm{Z}_j) = \frac{1}{A_n} \iint\limits_{\mathcal{S}} \mathrm{Z}_j(\frac{x}{R},\frac{y}{R}) \mathrm{H}_1(\frac{x-X_n}{r_n},\frac{y-Y_n}{r_n}) \mathrm{d}x\mathrm{d}y,
\label{eq:piston}
\end{equation}
\begin{equation}
t_n(\mathrm{Z}_j) = \frac{1}{A_n} \iint\limits_{\mathcal{S}} \mathrm{Z}_j(\frac{x}{R},\frac{y}{R}) \mathrm{H}_2(\frac{x-X_n}{r_n},\frac{y-Y_n}{r_n}) \mathrm{d}x\mathrm{d}y,
\label{eq:tip}
\end{equation}
\begin{equation}
T_n(\mathrm{Z}_j) = \frac{1}{A_n} \iint\limits_{\mathcal{S}} \mathrm{Z}_j(\frac{x}{R},\frac{y}{R}) \mathrm{H}_3(\frac{x-X_n}{r_n},\frac{y-Y_n}{r_n}) \mathrm{d}x\mathrm{d}y,
\label{eq:tilt}
\end{equation}
where $\mathcal{S}$ represents the surface of a single segment. The surface of an hexagon of radius $r$ is defined as $\mathcal{S} = \left\lbrace (x,y) \in \mathbb{R}^2 \vert x \in [X_n-r,X_n+r] , y \in [\zeta_1(x),\zeta_2(x)] \right\rbrace$ with
\[
    \zeta_1(x)= 
\begin{cases}
    Y_n-\sqrt{3}(X_n+x+r) & \text{for } -r < x-X_n < -\frac{r}{2}\\
    Y_n-\sqrt{3}r/2 & \text{for } -\frac{r}{2} < x-X_n < \frac{r}{2} \\
    Y_n+\sqrt{3}(X_n+x-r) & \text{for } \frac{r}{2} < x-X_n < r \\
\end{cases}
\]
and
\[
    \zeta_2(x)= 
\begin{cases}
    Y_n+\sqrt{3}(X_n+x+r) & \text{for } -r < x-X_n < -\frac{r}{2}\\
    Y_n+\sqrt{3}r/2 & \text{for } -\frac{r}{2} < x-X_n < \frac{r}{2} \\
    Y_n-\sqrt{3}(X_n+x-r) & \text{for } \frac{r}{2} < x-X_n < r \\
\end{cases}
\] and $X_n$ and $Y_n$ stand for the spatial coordinates of the center of the segment $n$. For instance, in the situation presented in Fig.~\ref{fig:segmented-pupil-variables}, $(X_1,Y_1) = (0,0)$ for the central segment; $(X_4,Y_4) = (-\Delta_x,-\Delta_y/2)$ for the segment number $4$;  $(X_{18},Y_{18}) = (2\Delta_x,\Delta_y)$ for the segment number $18$.

The same approach is used to reproduce the hexagonal polynomials $\mathrm{H}_j$ on a segmented pupil. Substituting $\mathrm{Z}_j$ by $\mathrm{H}_j$ in Equations~\ref{eq:piston}-\ref{eq:tilt}, we obtain the values for piston, tip, and tilt to apply to each segment. The coefficients are expressed as:
\begin{equation}
p_n(\mathrm{H}_j) = \frac{1}{A_n} \iint\limits_{\mathcal{S}} \mathrm{H}_j(\frac{x}{R},\frac{y}{R}) \mathrm{H}_1(\frac{x-X_n}{r_n},\frac{y-Y_n}{r_n}) \mathrm{d}x\mathrm{d}y,
\label{eq:piston_H}
\end{equation}
\begin{equation}
t_n(\mathrm{H}_j) = \frac{1}{A_n} \iint\limits_{\mathcal{S}} \mathrm{H}_j(\frac{x}{R},\frac{y}{R}) \mathrm{H}_2(\frac{x-X_n}{r_n},\frac{y-Y_n}{r_n}) \mathrm{d}x\mathrm{d}y,
\label{eq:tip_H}
\end{equation}
\begin{equation}
T_n(\mathrm{H}_j) = \frac{1}{A_n} \iint\limits_{\mathcal{S}} \mathrm{H}_j(\frac{x}{R},\frac{y}{R}) \mathrm{H}_3(\frac{x-X_n}{r_n},\frac{y-Y_n}{r_n}) \mathrm{d}x\mathrm{d}y.
\label{eq:tilt_H}
\end{equation}

For the sake of generality, we take into account a possible rotation of the segment $n$ by an angle $\theta_n$ around its center. This is included into the previous equations by rotating the modes around the central point of the segment $n$ denoted $(X_n,Y_n)$.
This implies the following change of variable in Equations~\ref{eq:piston}-\ref{eq:tilt_H} :
\begin{equation}
    \begin{split}
    \frac{x}{R} \rightarrow \frac{X_n + (x-X_n)\cos\theta_n -(y-Y_n)\sin\theta_n}{R}, \\
    \frac{y}{R} \rightarrow \frac{Y_n + (x-X_n)\sin\theta_n +(y-Y_n)\cos\theta_n}{R}.
    \end{split}
\end{equation}
The resultant formul\ae~for piston ($p_n$), tip ($t_n$) and tilt ($T_n$) decomposition are listed in Table~\ref{tab:coefficients-zernike_rotation} for the Zernike polynomials, and in Table~\ref{tab:coefficients-hexagonal_rotation} for the hexagonal polynomials.
We note that these formul\ae~only depend on the position of the segment, its radius $r_n$, the radius of the mode $R$ and on the angle $\theta_n$. Thus, once analytically calculated, they can be used for any configuration of segmented pupil (e.g. exotic pupil configurations like uneven ($X_n,Y_n$) positions, changing in the segment size $r_n$ across the pupil or different orientation $\theta_n$ of the segments).

\begin{table}[!htbp]
\centering
\caption{\bf Piston, tip, and tilt coefficients on the segment \boldmath$n$ for the decomposition of the $\mathbf{10}$ first Zernike polynomials \boldmath$\mathrm{Z}_j$ over a segmented aperture including an arbitrary rotation of the segment by $\theta_n$. }
\tiny
\begin{tabular*}{\textwidth}{l@{\extracolsep{\fill}} lll}
\hline
$j$ & $p_n(\mathrm{Z}_j)$ & $t_n(\mathrm{Z}_j)$ & $T_n(\mathrm{Z}_j)$  \\
\hline
$1$ & $1$ & $0$ & $0$  \\

 $2$ & $\frac{2}{R}X_n$ & $\frac{r_n}{R}\sqrt{\frac{5}{6}}\cos\theta_n$ & $-\frac{r_n}{R}\sqrt{\frac{5}{6}}\sin\theta_n$  \\

 $3$ & $\frac{2}{R} Y_n$ & $\frac{r_n}{R}\sqrt{\frac{5}{6}}\sin\theta_n$ & $\frac{r_n}{R}\sqrt{\frac{5}{6}}\cos\theta_n$  \\

 $4$ & $\frac{\sqrt{3}}{6R^2} (12X_n^2+12Y_n^2+5r_n^2-6R^2)$ & $\frac{r_n\sqrt{10}}{R^2} (X_n\cos\theta_n+Y_n\sin\theta_n)$ & $\frac{r_n\sqrt{10}}{R^2} (-X_n\sin\theta_n+Y_n\cos\theta_n)$  \\

 $5$ & $\frac{2\sqrt{6}}{R^2} X_n Y_n$ & $\frac{r_n\sqrt{5}}{R^2}(X_n\sin\theta_n+Y_n\cos\theta_n)$ & $\frac{r_n\sqrt{5}}{R^2}(X_n\cos\theta_n-Y_n\sin\theta_n)$  \\

 $6$ & $\frac{\sqrt{6}}{R^2}(X_n^2-Y_n^2)$ & $\frac{r_n\sqrt{5}}{R^2}(X_n\cos\theta_n-Y_n\sin\theta_n)$ & $-\frac{r_n\sqrt{5}}{R^2}(X_n\sin\theta_n+Y_n\cos\theta_n)$  \\

 $7$ &$\frac{\sqrt{2} Y_n}{R^3} \left(6 X_n^2+6 Y_n^2+5r_n^2-4R^2\right)$ & $\frac{r_n \sqrt{15}}{75 R^3} \left(150 X_n Y_n \cos\theta_n +...\right.$ & $\frac{r_n \sqrt{15}}{75 R^3} \left(-150 X_n Y_n \sin\theta_n +...\right.$ \\
  & 
  & \multicolumn{1}{r}{$\left.(75X_n^2 + 225Y_n^2 +42r_n^2-50R^2)\sin\theta_n\right)$} 
  & \multicolumn{1}{r}{$\left. (75X_n^2 + 225Y_n^2 +42r_n^2-50R^2)\cos\theta_n\right)$} \\

 $8$ &$\frac{\sqrt{2} X_n}{R^3} \left(6 X_n^2+6 Y_n^2+5r_n^2-4R^2\right)$ & $\frac{r_n \sqrt{15}}{75 R^3} \left(150 X_n Y_n \sin\theta_n +...\right.$ & $\frac{r_n \sqrt{15}}{75 R^3} \left(150 X_n Y_n \cos\theta_n -...\right.$ \\
  & 
  & \multicolumn{1}{r}{$\left.(225X_n^2 + 75Y_n^2 +42r_n^2-50R^2)\cos\theta_n\right)$} 
  & \multicolumn{1}{r}{$\left. (225X_n^2 + 75Y_n^2 +42r_n^2-50R^2)\sin\theta_n\right)$} \\

 $9$ &$ \frac{2 \sqrt{2}}{R^3} Y_n \left(3 X_n^2-Y_n^2\right)$ & $\frac{r_n\sqrt{15}}{R^3} \left(2X_nY_n\cos\theta_n+(X_n^2-Y_n^2)\sin\theta_n\right) $& $\frac{r_n\sqrt{15}}{R^3} \left(-2X_nY_n\sin\theta_n+(X_n^2-Y_n^2)\cos\theta_n\right) $ \\

 $10$ &$ \frac{2 \sqrt{2}}{R^3} X_n \left(X_n^2-3Y_n^2\right)$ & $\frac{r_n\sqrt{15}}{R^3} \left(-2X_nY_n\sin\theta_n+(X_n^2-Y_n^2)\cos\theta_n\right) $& $\frac{r_n\sqrt{15}}{R^3} \left(-2X_nY_n\cos\theta_n-(X_n^2-Y_n^2)\sin\theta_n\right) $ \\
\hline
\end{tabular*}
  \label{tab:coefficients-zernike_rotation}
\end{table}

\begin{table*}[!htbp]
\centering
\caption{\bf Piston, tip, and tilt coefficients on the segment \boldmath$n$ for the decomposition of the $\mathbf{10}$ first hexagonal polynomials \boldmath$\mathrm{H}_j$ over a segmented aperture. }
\tiny
\begin{tabular*}{\textwidth}{l@{\extracolsep{\fill}} lll}
\hline
$j$ & $p_n(\mathrm{H}_j)$ & $t_n(\mathrm{H}_j)$ & $T_n(\mathrm{H}_j)$  \\
\hline
$1$ & $1$ & $0$ & $0$  \\

$2$ & $\frac{2}{R}\sqrt{\frac{6}{5}} X_n$ & $\frac{r}{R}\cos\theta_n$ & $-\frac{r}{R}\sin\theta_n$ \\

$3$ & $\frac{2}{R} \sqrt{\frac{6}{5}} Y_n$ & $\frac{r}{R}\sin\theta_n$ & $\frac{r}{R}\cos\theta_n$ \\

$4$ 
& $\frac{1}{R^2}\sqrt{\frac{5}{43}} \left(5 r^2-5 R^2+12 \left(X_n^2+Y_n^2\right)\right)$
& $\frac{10r}{R^2} \sqrt{\frac{6}{43}} (X_n\cos\theta_n+Y_n\sin\theta_n)$ 
& $\frac{10r}{R^2} \sqrt{\frac{6}{43}} (-X_n\sin\theta_n+Y_n\cos\theta_n)$  \\

$5$ 
& $\frac{4}{R^2} \sqrt{\frac{15}{7}} X_n Y_n$ 
& $\frac{5r}{R^2} \sqrt{\frac{2}{7}} (X_n\sin\theta_n+Y_n\cos\theta_n)$ 
& $\frac{5r}{R^2} \sqrt{\frac{2}{7}} (X_n\cos\theta_n-Y_n\sin\theta_n)$ \\

$6$ 
& $\frac{2}{R^2} \sqrt{\frac{15}{7}} (X_n^2-Y_n^2)$
& $\frac{5r}{R^2} \sqrt{\frac{2}{7}} (X_n\cos\theta_n-Y_n\sin\theta_n)$ 
& $-\frac{5r}{R^2} \sqrt{\frac{2}{7}} (X_n\sin\theta_n+Y_n\cos\theta_n)$ \\

$7$ 
& $\frac{2}{R^3} \sqrt{\frac{14}{11055}} Y_n \left(125 r^2-84 R^2+...\right.$ 
& $\frac{2r}{R^3} \sqrt{\frac{7}{737}} \left(50X_n Y_n\cos\theta_n + ...\right.$ 
& $\frac{2r}{R^3} \sqrt{\frac{7}{737}} \left( -50X_n Y_n \sin\theta_n +...\right.$ \\
& \multicolumn{1}{r}{$\left. 150 \left(X_n^2+Y_n^2\right)\right)$}
& \multicolumn{1}{r}{$\left. (25X_n^2+75Y_n^2+14r^2-14R^2)\sin\theta_n\right)$ }
& \multicolumn{1}{r}{$\left. \left(14 r^2-14 R^2+25 X_n^2+75 Y_n^2\right)\cos\theta_n \right)$} \\

$8$ 
& $\frac{2}{R^3} \sqrt{\frac{14}{11055}} X_n \left(125 r^2-84 R^2+...\right.$ 
& $\frac{2r}{R^3} \sqrt{\frac{7}{737}} \left(50X_n Y_n\sin\theta_n + ... \right.$ 
& $\frac{2r}{R^3} \sqrt{\frac{7}{737}} \left(50X_n Y_n \cos\theta_n -...\right.$ \\
& \multicolumn{1}{r}{$\left. 150 \left(X_n^2+Y_n^2\right)\right)$}
& \multicolumn{1}{r}{$\left. (75X_n^2+25Y_n^2+14r^2-14R^2)\cos\theta_n\right)$ }
& \multicolumn{1}{r}{$\left. \left(14 r^2-14 R^2+75 X_n^2+25 Y_n^2\right)\sin\theta_n \right)$} \\

$9$ & $-\frac{4\sqrt{10}}{3 R^3}  Y_n \left(Y_n^2-3 X_n^2\right)$ 
& $\frac{10 \sqrt{3}r}{3R^3} \left(2X_n Y_n \cos\theta_n + ...\right.$ 
& $\frac{10 \sqrt{3} r}{ 3R^3} \left( -2X_n Y_n \sin\theta_n + ...\right.$ \\
&
& \multicolumn{1}{r}{$\left. (X_n^2-Y_n^2)\sin\theta_n \right)$} 
& \multicolumn{1}{r}{$\left. (X_n^2-Y_n^2)\cos\theta_n \right)$} \\

$10$ & $\frac{4}{R^3} \sqrt{\frac{70}{103}} X_n \left(X_n^2-3 Y_n^2\right)$ 
& $\frac{10r}{R^3} \sqrt{\frac{21}{103}} \left( -2X_n Y_n \sin\theta_n + ...\right.$ 
& $-\frac{10r}{R^3} \sqrt{\frac{21}{103}} \left(2X_n Y_n \cos\theta_n + ...\right.$  \\
&
& \multicolumn{1}{r}{$\left. (X_n^2-Y_n^2)\cos\theta_n \right)$} 
& \multicolumn{1}{r}{$\left. (X_n^2-Y_n^2)\sin\theta_n \right)$} \\

\hline
\end{tabular*}
  \label{tab:coefficients-hexagonal_rotation}
\end{table*}

\section{Evaluation and discussion}
\label{sec:fitting_error}


In this section, we discuss a standard error term affecting both numerical methods and our analytical approach: the fitting error. In this context, the performance our method yields over numerical approaches is examine. Finally, we outline some applications that would benefit from a rigorous analytical expression of the Zernike and hexagonal decomposition over an hexagonal pupil.

\subsection{Fitting error: analytical decomposition}
Due to the segmentation and the finite number of degrees of freedom, the reproduction of Zernike or hexagonal modes on a segmented pupil is subject to fitting errors.
\begin{figure}[!bp]
\centering
\includegraphics[width=7cm]{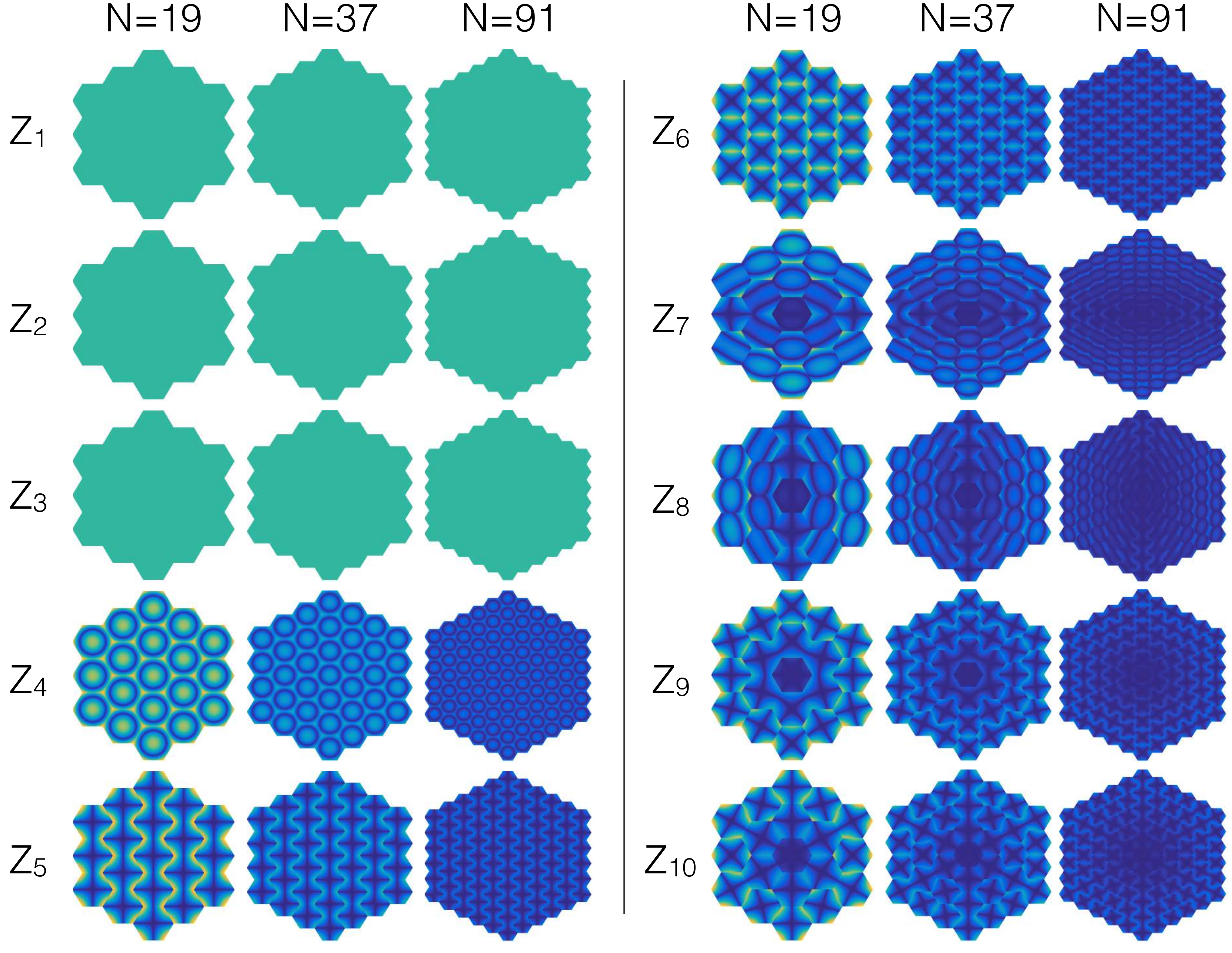}
\caption{Representation of the absolute errors $\lvert \varepsilon_n(\mathcal{Z}_j) \rvert$ obtained by means of simulations for the first ten Zernike modes. Three different pupils are considered ($N=19$, $37$ and $91$).}
\label{fig:fitted-zernike-modes}
\end{figure}
We propose in this section to calculate analytically the fitting error of a given configuration (i.e. with given parameters $r_n$, $\theta_n$ $R$ and $M$). We first define the fitting error at the position $(x,y)$ on the pupil (located on segment $n$) as the difference between the fitted and the theoretical expressions, first for the Zernike modes as:
\begin{equation}
\label{eq:epsilonZ}
\begin{aligned}
\varepsilon_n (x,y,\mathrm{Z}_j) = \Big[ p_n(\mathrm{Z}_j) \times \mathrm{H}_1\Big(\frac{x-X_n}{r_n},\frac{y-Y_n}{r_n}&\Big) \\
+ t_n(\mathrm{Z}_j) \times \mathrm{H}_2\Big(\frac{x-X_n}{r_n},\frac{y-Y_n}{r_n}&\Big) \\
+ T_n(\mathrm{Z}_j) \times \mathrm{H}_3\Big(\frac{x-X_n}{r_n},\frac{y-Y_n}{r_n}&\Big) \Big] \\
- \mathrm{Z}_j \Big(\frac{X_n + (x-X_n)\cos\theta_n -(y-Y_n)\sin\theta_n}{R}&,\\
\frac{Y_n + (x-X_n)\sin\theta_n +(y-Y_n)\cos\theta_n}{R}&\Big),
\end{aligned}
\end{equation}
and the fitting error for the hexagonal polynomials as:
\begin{equation}
\label{eq:epsilonH}
\begin{aligned}
\varepsilon_n (x,y,\mathrm{H}_j) = \Big[ p_n(\mathrm{H}_j) \times \mathrm{H}_1\Big(\frac{x-X_n}{r_n},\frac{y-Y_n}{r_n}&\Big) \\
+ t_n(\mathrm{H}_j) \times \mathrm{H}_2\Big(\frac{x-X_n}{r_n},\frac{y-Y_n}{r_n}&\Big) \\
+ T_n(\mathrm{H}_j) \times \mathrm{H}_3\Big(\frac{x-X_n}{r_n},\frac{y-Y_n}{r_n}&\Big) \Big] \\
- \mathrm{H}_j \Big(\frac{X_n + (x-X_n)\cos\theta_n -(y-Y_n)\sin\theta_n}{R}&,\\
\frac{Y_n + (x-X_n)\sin\theta_n +(y-Y_n)\cos\theta_n}{R}&\Big).
\end{aligned}
\end{equation}
The absolute value of the fitting errors $\left\lvert \varepsilon \right\rvert$ for the ten first Zernike modes are visible in Fig.~\ref{fig:fitted-zernike-modes}. 
As for now, we consider a regular pupil, i.e. $(X_n,Y_n)$ distributed according to a regular hexagonal grid, with $\theta_n = 0$ and $r_n$ constants for all the segments. Finally, for the Zernike modes we introduce the mean squared error on the segment $n$ that depends on $n$ and on the index $j$:

\begin{equation}
\begin{aligned}
\overline{\varepsilon_n^2 (\mathrm{Z}_j)} = \frac{1}{A_n} \iint\limits_{\mathcal{S}} \varepsilon_n^2 (x,y,\mathrm{Z}_j) \mathrm{d}x \mathrm{d}y,
\end{aligned}
\end{equation}
and similarly for the hexagonal modes:
\begin{equation}
\begin{aligned}
\overline{\varepsilon_n^2 (\mathrm{H}_j)} = \frac{1}{A_n} \iint\limits_{\mathcal{S}} \varepsilon_n^2 (x,y,\mathrm{H}_j) \mathrm{d}x \mathrm{d}y.
\end{aligned}
\end{equation}

The results of these two integrals are given in Table~\ref{tab:fitting_error}. As expected, the fitting error for piston, tip and tilt modes is null, meaning that a segmented mirror can perfectly reproduce one of them.
For modes $4$ to $6$ the error is the same for all the segments and only depends on the ratio $r_n/R$. This effect is clearly visible in Fig.~\ref{fig:fitted-zernike-modes}, where the error figure for modes 4 to 6 is independent from the considered segment.
For higher order modes, terms depending on the positions of the segments $X_n$ and $Y_n$ appear in the equations and the error figure in Fig.~\ref{fig:fitted-zernike-modes} is now different from one segment to another. Finally, for modes $9$ and $10$, a term in $\cos 6 \theta_n$ appears in the equations meaning that the error figure now depends on the rotation of the segment 
because the system has a $\pi/3$ symmetry.

\begin{table}[!b]
\centering
\caption{\bf Mean fitting error values over the segment \boldmath$n$ for the $\mathbf{10}$ first Zernike and hexagonal modes.}
\begin{tabular*}{\linewidth}{l @{\extracolsep{\fill}} l l}
\hline
$j$ & $\overline{\varepsilon_n^2(\mathrm{Z}_j)}$ for $\mathrm{Z}_j$ modes & $\overline{\varepsilon_n^2(\mathrm{H}_j)}$ for $\mathrm{H}_j$ modes \\
\hline
$1$ & $0$ & $0$ \\

$2$ & $0$ & $0$ \\

$3$ & $0$ & $0$ \\

$4$ & $\frac{43}{60}\left(\frac{r}{R}\right)^4$ & $\left(\frac{r}{R}\right)^4$\\

$5$ & $\frac{7}{10}\left(\frac{r}{R}\right)^4$ & $\left(\frac{r}{R}\right)^4$\\

$6$ & $\frac{7}{10}\left(\frac{r}{R}\right)^4$ & $\left(\frac{r}{R}\right)^4$\\

$7$ 
& $\frac{1400 r^4\left(21 X_n^2+64 Y_n^2\right)+2211 r^6}{3500 R^6}$ 
& $\frac{1400r^4 \left(21 X_n^2+64 Y_n^2\right)+2211 r^2}{2211 R^6}$ \\

$8$ 
& $\frac{1400r^4 \left(64 X_n^2+21 Y_n^2\right)+2211 r^6}{3500 R^6}$ 
& $\frac{1400r^4 \left(64 X_n^2+21 Y_n^2\right)+2211 r^6}{2211 R^6}$\\

$9$
& $\frac{1176r^4 \left(X_n^2+Y_n^2\right) + r^6\left(83 - 20\cos6\theta_n\right)}{140 R^6}$ 
& $\frac{1176r^4\left(X_n^2+Y_n^2\right) + r^6\left(83 - 20\cos6\theta_n\right)}{63 R^6}$ \\

$10$ 
& $\frac{1176r^4 \left(X_n^2+Y_n^2\right) + r^6\left(83 + 20\cos6\theta_n\right)}{140 R^6}$  
& $\frac{1176r^4 \left(X_n^2+Y_n^2\right) + r^6\left(83 + 20\cos6\theta_n\right)}{103 R^6}$ \\

\hline
\end{tabular*}
  \label{tab:fitting_error}
\end{table}

\begin{figure*}[!tbp]
\centering
\includegraphics[width=.95\linewidth]{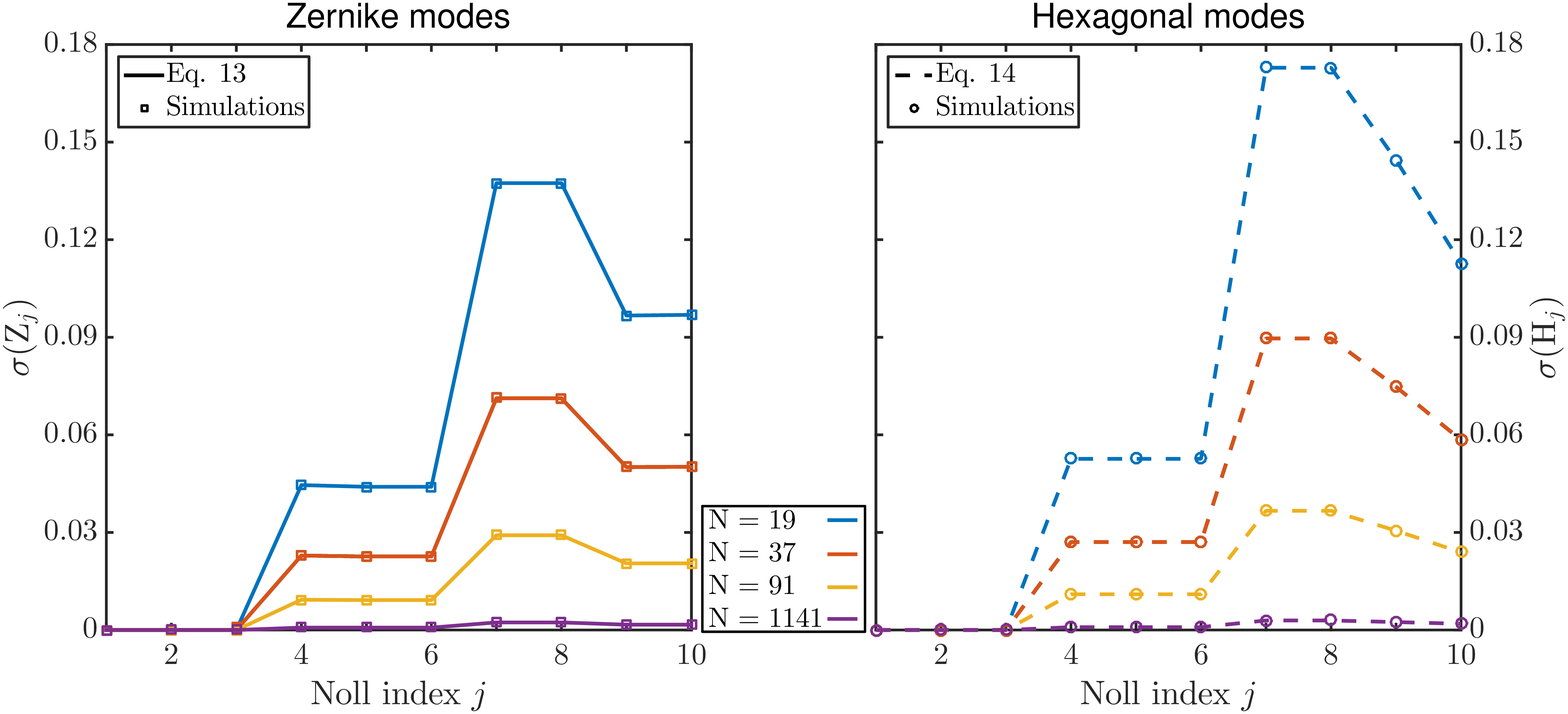}
\caption{Representation of the total fitting error for the first 10 Zernike polynomials (left) and hexagonal polynomials (right) for different numbers of segments comprising the mirror. The dots represent the values obtained by means of numerical simulations while the lines represent the theoretical values given in Table~\ref{tab:fitting_error}.}
\label{fig:fitting_error}
\end{figure*}

To further confirm these analytical expressions, we evaluate the total error over the segmented pupil for a given mode obtained with numerical simulations by creating the corresponding mirror configuration using the formul\ae~from Tables~\ref{tab:coefficients-zernike_rotation} or \ref{tab:coefficients-hexagonal_rotation}, and by comparing to that of the theoretical value. When considering the Zernike modes, the total error is given by:
\begin{equation}
\label{eq:sigma_Z}
    \sigma({\mathrm{Z}_j})= \sqrt{\frac{A}{\pi R^2} \times  \sum_n \overline{\varepsilon_n^2(\mathrm{Z}_j)}} = \sqrt{\frac{3\sqrt{3}r^2}{2\pi R^2} \times  \sum_n \overline{\varepsilon_n^2(\mathrm{Z}_j)}},
\end{equation}
while considering the hexagonal modes we have:
\begin{equation}
\label{eq:sigma_H}
    \sigma({\mathrm{H}_j})= \sqrt{\frac{2A}{3\sqrt{3} R^2} \times  \sum_n \overline{\varepsilon_n^2(\mathrm{H}_j)}} = \sqrt{\frac{r^2}{R^2} \times  \sum_n \overline{\varepsilon_n^2(\mathrm{H}_j)}},
\end{equation}
where the terms appearing before the sums in Equation~\ref{eq:sigma_Z} and Equation \ref{eq:sigma_H} are the ratios between the area of one segment and the normalization coefficients of the Zernike or hexagonal polynomials. The comparison between the theoretical fitting errors and the values obtained by means of simulations are presented in Fig.~\ref{fig:fitting_error}, for the Zernike polynomials (left) and for the hexagonal polynomials (right).
The error has been calculated for four different configurations, with $N=19$, $37$, $91$, and $1141$. As expected, and as it is observable in Fig.~\ref{fig:fitted-zernike-modes} and Fig.~\ref{fig:fitting_error}, the larger the number of segments comprising the mirror, the lower the fitting error. The match between simulated fitting errors and analytical predicted expressions is fairly found. 
For the Zernike modes (Fig.~\ref{fig:fitting_error}, left), various plateaux are observable. While the reason for the first plateau occurring from the Noll index 1 to 3 is trivial, the three others originate from the relation that exist between the pairs of modes $5$ and $6$, $7$ and $8$, and $9$ and $10$ because of the rotation of a centro-symetric polynomials around its center. We note that the correspondence between the mode $4$ and the modes $5$ and $6$ is only fortuitous. Basically, the same trend can be seen in Fig. \ref{fig:fitting_error} (right) for the hexagonal modes, and for the same reasons, except for numbers $9$ and $10$, where the $45$\degre~rotation of the non centro-symetric polynomials provides different results.
The very good correspondence between the simulated errors and their expected theoretical values further confirm the validity of the proposed analytical decomposition.


\subsection{Fitting error: analytical vs. numerical decompositions}

\begin{figure*}[!b]
\centering
\includegraphics[height=.447\linewidth]{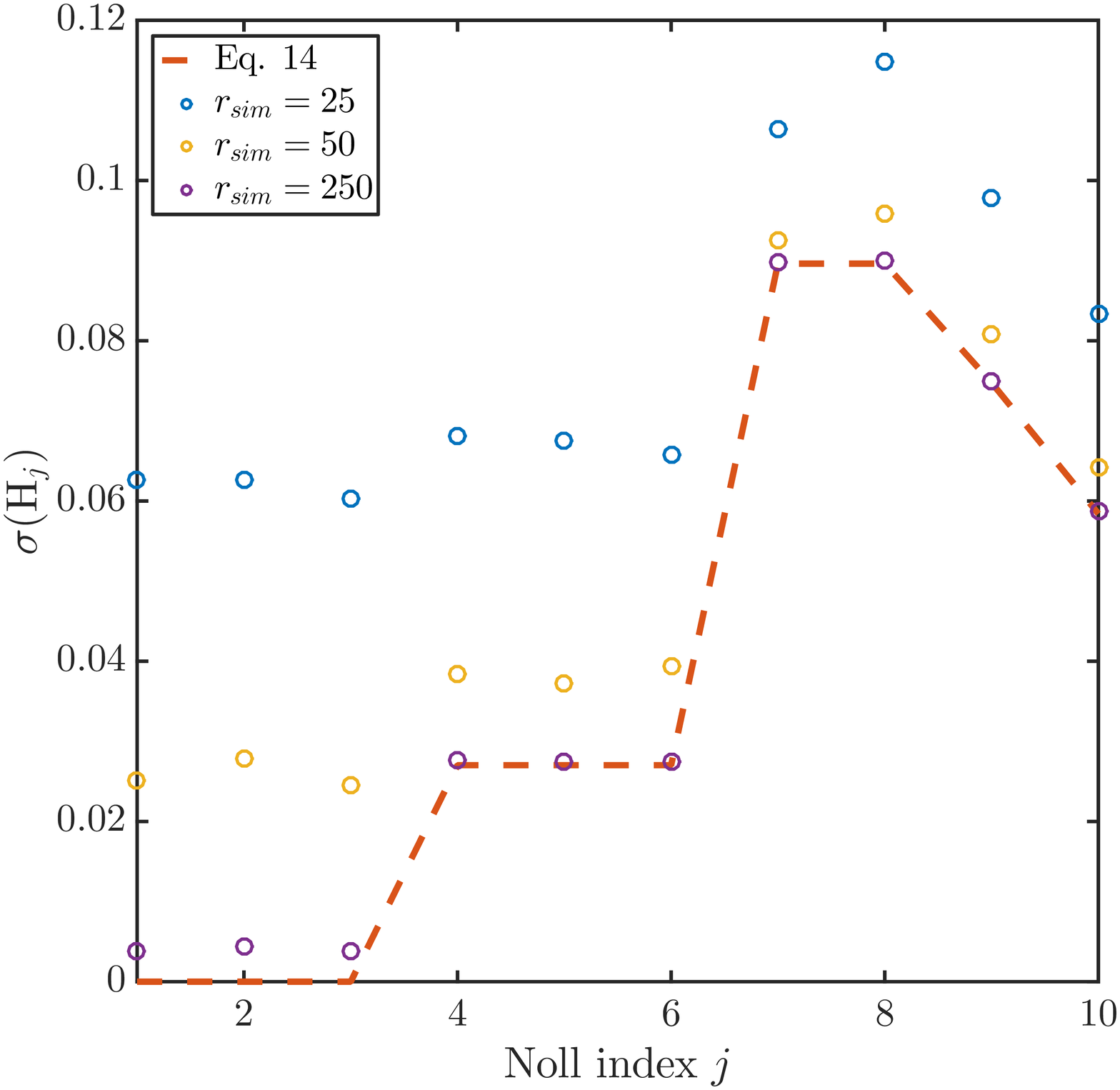}
\hfill
\includegraphics[height=.45\linewidth]{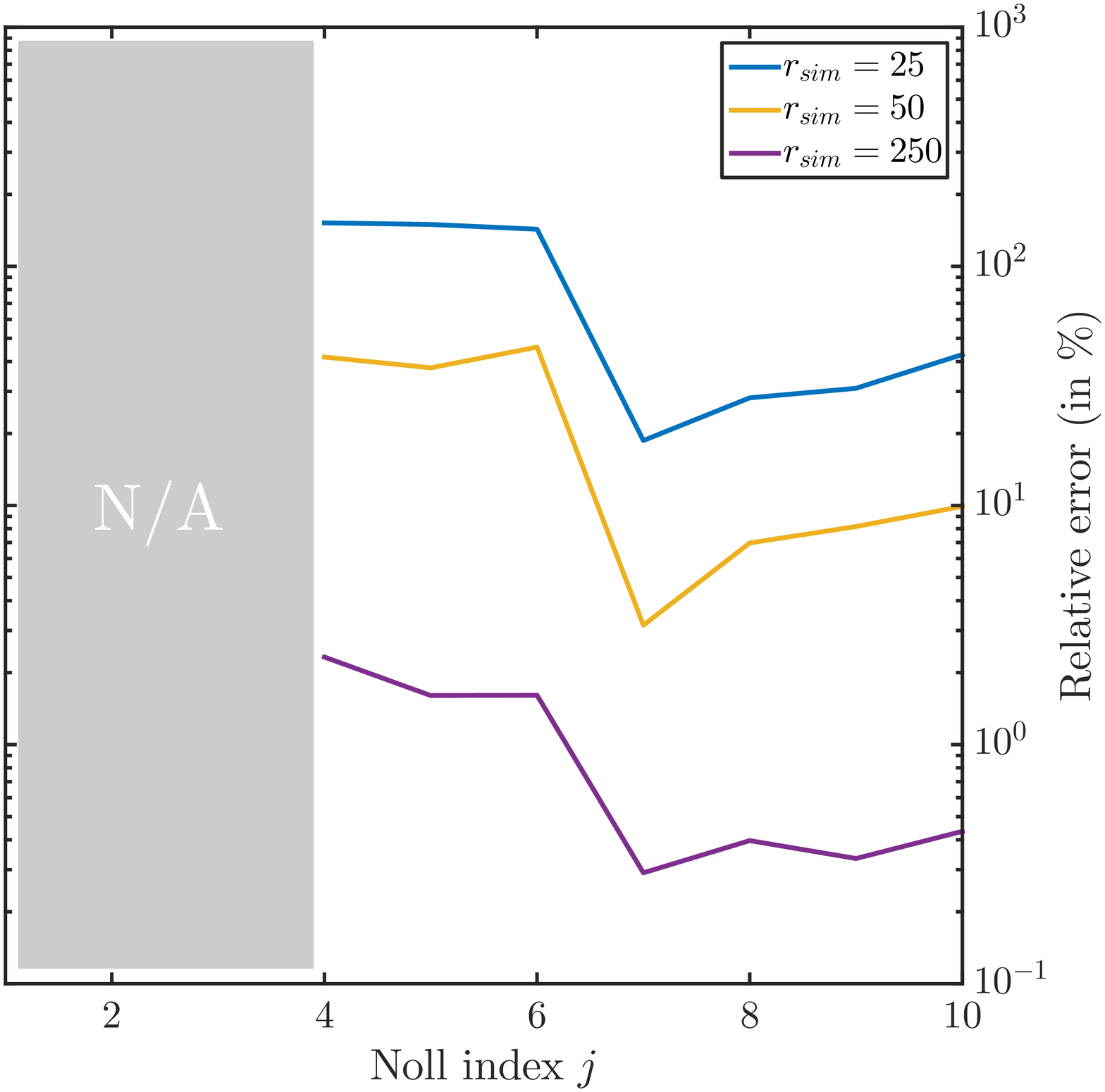}
\caption{Representation of the total fitting error $\sigma(\mathcal{H}_j)$ (left) and its relative error with regard to Eq.~\ref{eq:sigma_H} (right) for the first 10 hexagonal polynomials for different values of $r_{sim}$ used to numerically compute the piston, tip and tilt coefficients on each segment. The pupil used for this simulation is composed of $N=37$ segments.}
\label{fig:numerical_fitting_error}
\end{figure*}

In this section, we discuss the performance our method yields over a numerical approach by comparing the fitting error obtained with these two methods. For the numerical method we use the following charts of simulation steps: 
\begin{enumerate}
\item Calculation of the numerical decomposition coefficients
\begin{enumerate}
\item We simulate a unitary discrete pupil with the desired number of segments $N$. Each segment is given a radius in pixel denoted $r_{sim}$.
\item We apply the desired polynomial (Zernike or hexagonal) onto the pupil.
\item We retrieve each segments comprising the pupil and project their respective signal onto a piston, tip and tilt basis to obtain the decomposition coefficients.
\end{enumerate}
\item Fitting error estimation
\begin{enumerate}
\item We apply these coefficients to the segments of a new pupil and obtain the fitted hexagonal mode. The radius of these new segments are chosen to be large enough to best avoid the errors introduced by the numerical discretization when calculating the fitting errors.
\item We numerically compute the difference between the fitted mode and the theoretical one, as expressed in Equation~\ref{eq:epsilonH}.
\end{enumerate}
\end{enumerate}

This process is then repeated for each mode. We present in Fig.~\ref{fig:numerical_fitting_error} (left) the total error $\sigma(\mathrm{H}_j)$ obtained with different values of $r_{sim}$ for the first 10 hexagonal polynomials and in Fig.~\ref{fig:numerical_fitting_error} (right) the corresponding relative error between $\sigma(\mathrm{H}_j)$ and the theoretical fitting error obtained with Eq.~\ref{eq:sigma_H}. The first conclusion is that the smaller the radius $r_{sim}$ of the segment, the greater the error $\sigma_{\mathrm{H}_j}$. This is directly due to the discretization of the segments in a finite number of pixels that leads to a wrong estimation of the coefficients. For a value $r_{sim}=25$,  the relative error can reach up to $100\%$ error for the modes $4$, $5$ and $6$ which is equivalent to a doubling of the theoretical fitting error. With a bigger value of $r_{sim}=250$, this relative error scales down to $1\%$, which can still not be considered as negligible.


What is also distinguishable in Fig.~\ref{fig:numerical_fitting_error} (left) and not intuitive, is that the numerical method can not perfectly reproduce the first mode $\mathrm{H}_{1}$. This is due to the fact that simulated segments are not exactly symmetric with respect to their respective center, as their edge has to be cropped following the boundary equations given in Note~3. This induce non-zero tip-tilt calculated coefficients while there is no tip-tilt on the segments, leading to a wrong estimation of the pupil configuration. 
Finally, the curves in Fig.~\ref{fig:numerical_fitting_error} (right) show two different regimes separated in two plateaux: one from modes $4$ to $6$ and another one from $7$ to $10$. This behavior related to the spatial frequencies we want to decompose in each case. The higher the spatial frequency (i.e. the radial order of the Zernike mode) the lower the importance of the signal at the border of the segment thus leading to a lower relative error.

\subsection{Discussion and applications}
The analytical method presented in this paper  allows a much faster Zernike and hexagonal decomposition than numerical methods. 
Numerical methods are inevitably subject to computational and memory costs that can be impressive in particular when segments require to be highly sampled or when segments are numerous. Our method avoids these fundamental limitations, as well as being totally independent of the pupil configuration.
Given any system parameters, the analytical method suffers from a marginal loss in accuracy due to the fitting error compared to that of numerical methods, and the relative error can be significant. Fitting errors affecting numerical methods can be limited by increasing the size of the segment, but it will consequently increase the computation time needed to calculate the coefficients which scales with $N\times r^2_{sim}$. 

The analytical method presented in this paper has many potential applications in particular in astronomical instrumentation with highly segmented telescopes (ELTs). For instance, our method could be used both as a complement and an extension of the work presented in~\cite{Leboulleux18}. In this study, an analytical error budget is proposed to infer the impact of the cophasing misalignment on the coronagraphic performances. The analytical model proposed in~\cite{Leboulleux18} offers the advantage of being low time consuming by avoiding long Monte-Carlo simulation. Complementing this analytical study by our analytical method for the Zernike decomposition is thus a must. So far, the segmented pupil only undergoes local independent piston on each segment. The study could thus benefit from the analytical coefficients of Zernike modes in order to increase the size of the decomposition basis by including global modes over the pupil.

\section{Conclusion}

We provide in this paper an analytical formulation of piston, tip, and tilt coefficients to each segment of a segmented pupil to fit either the Zernike modes or the hexagonal polynomials.
The decomposition obtained with traditional methods based on numerical calculations are dependent on (1) the pupil configuration and (2) the sampling used for each segment, and are computational time and memory consuming. In case (1), for each modification of the pupil, the numerical process has to be repeated to obtain the new coefficients. For case (2), in order to obtain precise results, the radius of a simulated segment $r_{sim}$ has to be large enough in order to limits the edge effects. The theoretical formul\ae~presented in this paper are based on the continuous limit of the process ($r_{sim} \rightarrow \infty$) and naturally offer the advantages of being mathematically exact and entirely independent from the number and positions of the segments comprising the mirror. They can therefore be used for analytical calculations dealing with segmented apertures. Even if mathematically exact, the analytical decomposition over a segmented pupil is subject to fitting errors, but with the lowest possible loss of accuracy compared to numerical methods.

As an example of application, the future of ground-based astronomy in the next decades is bound to the incoming generation of extremely large telescopes, especially for exoplanet direct detection, which is a major driver for present and future observing programs. ELTs represent a major change in dimension, wavefront control and execution time. In this context, phasing optics that correct for the misalignment of individual segments of the primary segmented mirror (798 individual segments of 1.4 m for the 39 m ELT) would benefit from analytical procedures in many areas including: control, phasing, monitoring, performance prediction and evaluation, and system optimization.





\section*{Funding}
P. Janin-Potiron is grateful to the French Aerospace Lab (ONERA) for supporting his postdoctoral fellowship.
This work has been partially supported by the LABEX FOCUS (grant DIR-PDC-2016-TF) and the VASCO research program at ONERA.

\bibliography{sample}

\end{document}